\documentclass[runningheads]{llncs}
\usepackage{graphicx}
\usepackage{multirow}
\usepackage{longtable}
\usepackage{dirtree}

\usepackage{todonotes}
\begin{document} 
\title{V3C -- a Research Video Collection}
\author{Luca Rossetto\inst{1} \and Heiko Schuldt\inst{1} \and George Awad\inst{2} \and Asad A. Butt\inst{2} 
}
\institute{
Databases and Information Systems Research Group \\ 
Department of Mathematics and Computer Science\\ University of Basel, Switzerland \\
\email{firstname.lastname@unibas.ch} \and
National Institute of Standards and Technology\\
Information Technology Laboratory \\
Information Access Division\\
Gaithersburg, MD, USA\\
\email{firstname.lastname@nist.gov}
}
\maketitle   
\begin{abstract}
With the widespread use of smartphones as recording devices and the massive growth in bandwidth, the number and volume of video collections has increased significantly in the last years. This poses novel challenges to the management of these large-scale video data and especially to the analysis of and retrieval from such video collections. At the same time, existing video datasets used for research and experimentation are either not large enough to represent current collections or do not reflect the properties of video commonly found on the Internet in terms of content, length, or resolution.

In this paper, we introduce the  \emph{Vimeo Creative Commons Collection}, in short V3C, a collection of 28'450 videos (with overall length of about 3'800 hours) published under creative commons license on Vimeo. V3C comes with a shot segmentation for each video, together with the resulting keyframes in original as well as reduced resolution and additional metadata. It is intended to be used from 2019 at the International large-scale TREC Video Retrieval Evaluation campaign (TRECVid).

%\keywords{First keyword  \and Second keyword \and Another keyword.}
\end{abstract}

\section{Introduction}

Over recent years, video has become a significant portion of the overall data which populates the web. This has been due to the fact that the production and distribution of video has shifted from a complex and costly endeavor to something accessible to everybody with a smart phone or similar device and a connection to the internet. This growth of content enabled new possibilities in various research areas which are able to make use of it. Despite the access to such large amounts of data, there remains a need for standardized datasets for computer vision and multimedia tasks. Multiple such datasets have been proposed over the years. A prominent example of a video dataset is the IACC~\cite{over2009creating} which has been used for several years now for international evaluation campaigns such as TRECVid~\cite{2017trecvidawad}. Other examples of datasets in the video context include the YFCC100M~\cite{thomee2016yfcc100m} which, despite being sourced from the photo-sharing platform Flickr\footnote{\url{https://flickr.com/}}, contains a considerable amount of video material, the Movie Memorability Database~\cite{cohendet2018annotating} which is comprised of memorable sequences from 100 Hollywood-quality movies or the YouTube-8M~\cite{abu2016youtube} dataset which in contrast, despite being sourced from YouTube\footnote{\url{https://youtube.com/}}, does not contain the original videos themselves. The content of all of these collections does, however, differ substantially from the type of web video commonly found `in the wild'~\cite{rossetto2017web}.

In this paper, we present the \emph{Vimeo Creative Commons Collection} or \emph{V3C} for short. It is composed of 28'450 videos collected from the video sharing platform Vimeo\footnote{\url{https://vimeo.com/}}. Apart from the videos themselves, the collection includes meta and shot-segmentation data for each video, together with the resulting keyframes in original as well as reduced resolution. The objective of V3C is to eventually complement or even replace existing collections in real-world video retrieval evaluation campaigns and thus to tailor the latter more to the type of video that can be found on the Internet.

The remainder of this paper is structured as follows: Section~\ref{sec:process} gives an overview of the process of how the collection was assembled and Section~\ref{sec:v3c} introduces the collection itself, its structure and some of its properties. Finally, Section~\ref{sec:conclusion} concludes.

\section{Collection Process}
\label{sec:process}

The requirements for usable video sources from which to compile a collection were as follows:
\begin{itemize}
\item The platform must be freely accessible.
\item It must host a large amount of diverse and contemporary video content.
\item At least a portion of the content must be published under a creative commons\footnote{\url{https://creativecommons.org/}} license and can therefore be redistributed in such a collection.
\end{itemize}

Two candidates for such collections are Vimeo and YouTube.
Vimeo was chosen over YouTube because while YouTube offers its users the possibility to publish videos under a creative commons attribution license which would allow the reuse and redistribution of the video material, YouTube's \emph{Terms of Service}~\cite{YoutubeTerms} explicitly forbid the download of any video on the platform for any reason other than playback in the context of a video stream.

We utilized the Vimeo categorization system for video collection. Videos are placed in 16 broad categories, which are further divided into subcategories. Videos in each category were examined to determine if they satisfied the `real world' requirements for the collection. Four top level categories were included in the collection, while 3 were excluded. For the remaining 9 categories, only some subcategories were included. The following are the 4 categories completely included in the collection:
`Personal', `Documentary', `Sports' and `Travel'.

\begin{figure}[t]
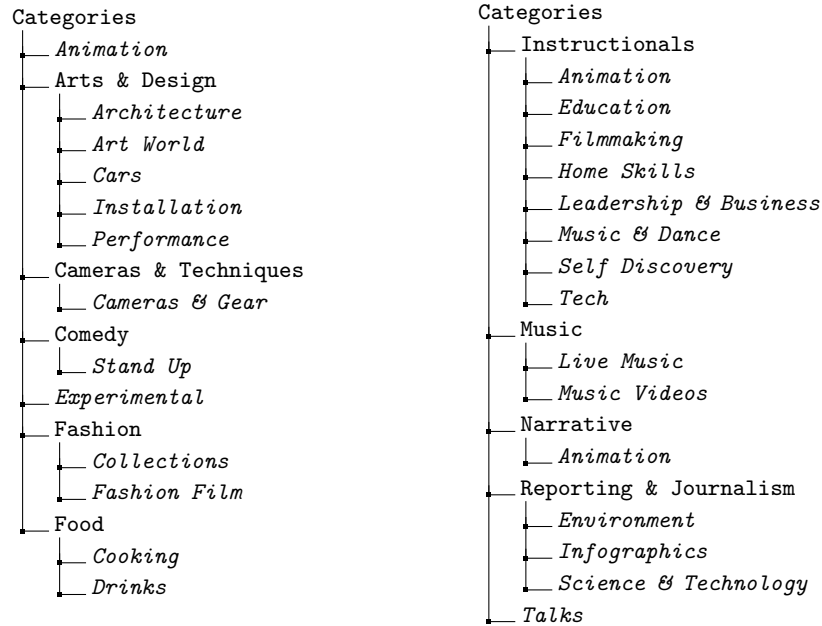


\begin{tabular}{ll}

\begin{minipage}{0.5\textwidth}
\dirtree{%
.1 Categories.
.2 \emph{Animation}.
.2 Arts \& Design.
.3 \emph{Architecture}.
.3 \emph{Art World}.
.3 \emph{Cars}.
.3 \emph{Installation}.
.3 \emph{Performance}.
.2 Cameras \& Techniques.
.3 \emph{Cameras \& Gear}.
.2 Comedy.
.3 \emph{Stand Up}.
.2 \emph{Experimental}.
.2 Fashion.
.3 \emph{Collections}.
.3 \emph{Fashion Film}.
.2 Food.
.3 \emph{Cooking}.
.3 \emph{Drinks}.
}
\vspace{3mm}
\end{minipage}
&
\begin{minipage}{0.5\textwidth}
\dirtree{%
.1 Categories.
.2 Instructionals.
.3 \emph{Animation}.
.3 \emph{Education}.
.3 \emph{Filmmaking}.
.3 \emph{Home Skills}.
.3 \emph{Leadership \& Business}.
.3 \emph{Music \& Dance}.
.3 \emph{Self Discovery}.
.3 \emph{Tech}.
.2 Music.
.3 \emph{Live Music}.
.3 \emph{Music Videos}.
.2 Narrative.
.3 \emph{Animation}.
.2 Reporting \& Journalism.
.3 \emph{Environment}.
.3 \emph{Infographics}.
.3 \emph{Science \& Technology}.
.2 \emph{Talks}.
}
\end{minipage}
\end{tabular}
	\caption{Removed categories and subcategories are \emph{emphasized}.}
	\label{fig:v3c-cat_removal}
\end{figure}

An overview of the excluded categories can be seen in Figure~\ref{fig:v3c-cat_removal}. Categories that had very low visual diversity (such as `Talks'), or did not represent real world scenarios were removed. Categories (or subcategories) with a lot of animation/graphics, or non standard content with little or no describable activity were excluded from the collection.  
Videos from the selected categories were then filtered by duration and license.

The obtained list of candidate videos was downloaded from Vimeo using an open-source video download utility\footnote{\url{https://github.com/rg3/youtube-dl}}. The download was performed sequentially in order to not cause unnecessary load on the side of the platform. All downloaded videos were subsequently checked to ensure they could be properly decoded by a commonly used video decoding utility\footnote{\url{https://ffmpeg.org/}}.

The videos were segmented and analyzed using the open-source content-based video retrieval engine Cineast~\cite{rossetto2014cineast}. Videos with a distribution of segment lengths which were sufficiently different from the mean were flagged for manual inspection as this indicated either very low or very high visual diversity as in the cases of either mostly static frames or very noisy videos. During this step, videos were also checked to ensure that the collection does not contain exact duplicates.

Out of the remaining videos, three subsets with increasing size were randomly selected. Sequential numerical ids were assigned to the selected videos in such a way that the first id in the second part is one larger than the last id in the first part and so on, in order to facilitate situations in which multiple parts are to be used in conjunction.

\section{The Vimeo Creative Commons Collection}
\label{sec:v3c}
The following provides an overview of the structure as well as various technical and semantic properties of the \emph{Vimeo Creative Commons Collection}.

\subsection{Collection Structure}

The collection consists of 28'450 videos with a duration between 3 and 60 minutes each and a total combined duration of slightly above 3'800 hours, divided into three partitions. Table \ref{tab:3:v3c-overview} provides an overview of the three partitions. Similar to the IACC, the V3C also includes a \emph{master shot reference} which segments every video into sequential non-overlapping parts, based on the visual content of the videos. For every one of these parts, a full resolution representative key-frame as well as a thumbnail image of reduced resolution is provided. Additionally, there are meta data files containing both technical as well as semantic information for every video which was also obtained from Vimeo.

\begin{table}[b]
	\centering
	\caption{Overview of the partitions of the V3C}
	\label{tab:3:v3c-overview}
    \setlength{\tabcolsep}{0.6em}
	\begin{tabular}{|l|c|c|c|c|}
		\hline
		Partition & \multicolumn{1}{c|}{V3C1} & \multicolumn{1}{c|}{V3C2} & \multicolumn{1}{c|}{V3C3} & \multicolumn{1}{c|}{Total} \\ \hline \hline
		File Size (videos) & 1.3TB & 1.6TB & 1.8TB & 4.8TB \\ \hline
        File Size (total) & 2.4TB & 3.0TB & 3.3TB & 8.7TB \\ \hline
		Number of Videos & 7'475 & 9'760 & 11'215 & 28'450 \\ \hline
		\begin{tabular}[c]{@{}l@{}}Combined\\ Video Duration\end{tabular} & \begin{tabular}[c]{@{}l@{}}1'000 hours, \\ 23 minutes, \\ 50 seconds\end{tabular} & \begin{tabular}[c]{@{}l@{}}1'300 hours, \\ 52 minutes,\\ 48 seconds\end{tabular} & \begin{tabular}[c]{@{}l@{}}1'500 hours, \\ 8 minutes, \\ 57 seconds\end{tabular} & \begin{tabular}[c]{@{}l@{}}3801 hours, \\ 25 minutes, \\ 35 seconds\end{tabular} \\ \hline
		Mean Video Duration & \begin{tabular}[c]{@{}l@{}}8 minutes, \\ 2 seconds\end{tabular} & \begin{tabular}[c]{@{}l@{}}7 minutes, \\ 59 seconds\end{tabular} & \begin{tabular}[c]{@{}l@{}}8 minutes, \\ 1 seconds\end{tabular} & \begin{tabular}[c]{@{}l@{}}8 minutes, \\ 1 seconds\end{tabular} \\ \hline
		Number of Segments & 1'082'659 & 1'425'454 & 1'635'580 & 4'143'693 \\ \hline
	\end{tabular}
\end{table}

\begin{figure}
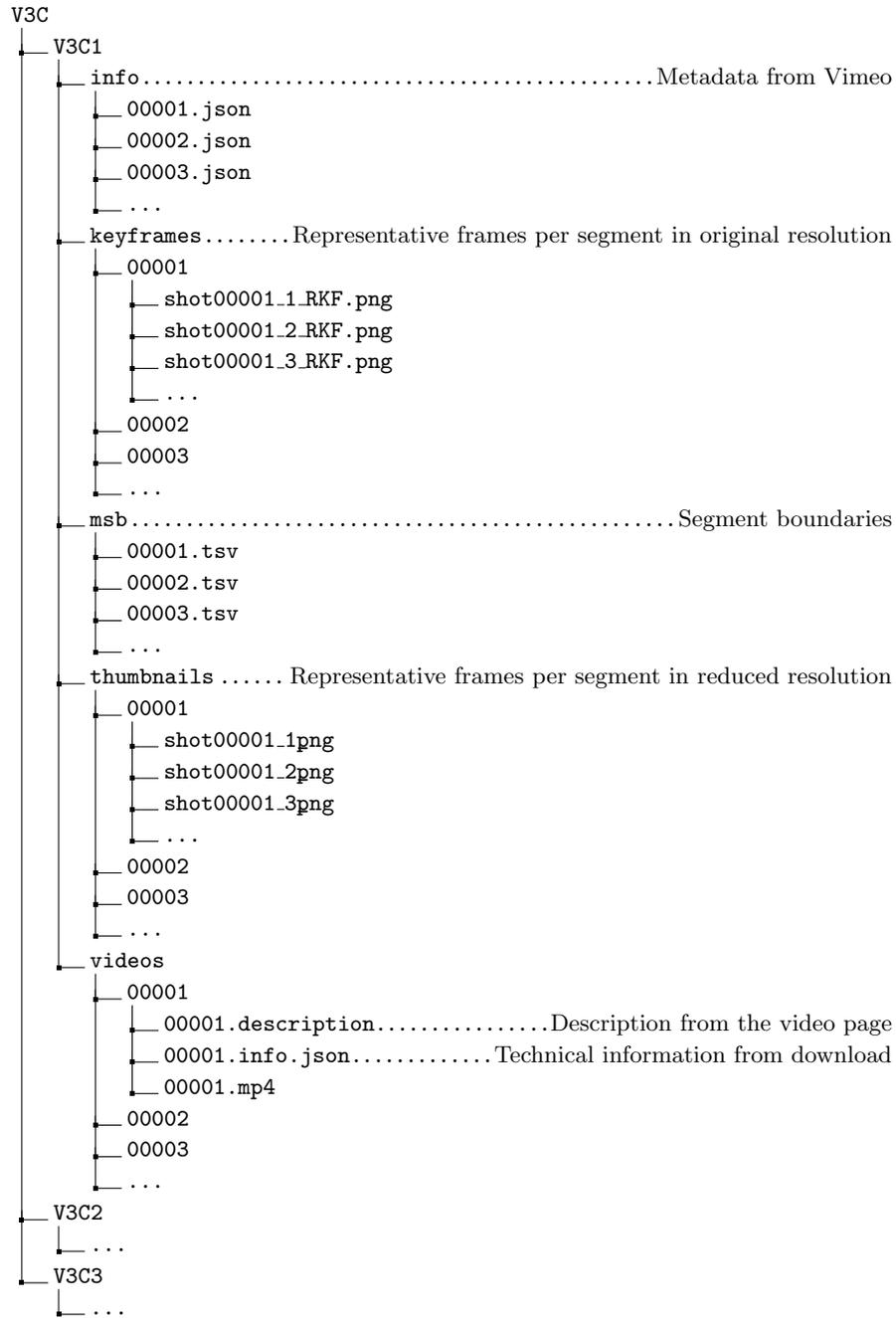

\dirtree{%
.1 V3C.
.2 V3C1.
.3 info\DTcomment{Metadata from Vimeo}.
.4 00001.json.
.4 00002.json.
.4 00003.json.
.4 \dots .
.3 keyframes\DTcomment{Representative frames per segment in original resolution}.
.4 00001.
.5 shot00001\_1\_RKF.png.
.5 shot00001\_2\_RKF.png.
.5 shot00001\_3\_RKF.png.
.5 \dots .
.4 00002.
.4 00003.
.4 \dots .
.3 msb\DTcomment{Segment boundaries}.
.4 00001.tsv.
.4 00002.tsv.
.4 00003.tsv.
.4 \dots .
.3 thumbnails\DTcomment{Representative frames per segment in reduced resolution}.
.4 00001.
.5 shot00001\_1\.png.
.5 shot00001\_2\.png.
.5 shot00001\_3\.png.
.5 \dots .
.4 00002.
.4 00003.
.4 \dots .
.3 videos.
.4 00001.
.5 00001.description\DTcomment{Description from the video page}.
.5 00001.info.json\DTcomment{Technical information from download}.
.5 00001.mp4 .
.4 00002.
.4 00003.
.4 \dots .
.2 V3C2.
.3 \dots .
.2 V3C3.
.3 \dots .
}
\caption{Directory structure of the V3C}
	\label{fig:3:dirtree}
\end{figure}

Every video in the collection has been assigned a sequential numerical id. These ids are then used for all aspects of the collection. Figure \ref{fig:3:dirtree} illustrates the directory structure which is used to organize the different aspects of the collection. This structure is identical for all three partitions. The \emph{info} directory contains one json-file per video which holds metadata obtained from Vimeo. This metadata contains both semantic information -- such as video title, description and associated tags -- as well as technical information including video duration, resolution, license and upload date. The \emph{msb} directory contains for each video a file in tab-separated format which lists the temporal start and end-positions for every automatically detected segment in a video. The \emph{keyframes} and \emph{thumbnails} directories each contain a subdirectory per video which hold one representative frame per video segment in a PNG format. The \emph{keyframes} are kept in the original video resolution while the thumbnails are downscaled to a width of 200 pixels. Finally the \emph{videos} directory contains a subdirectory per video, each of which containing the video itself as well as the video description and a file with technical information describing the download process.

\subsection{Statistical Properties}

The following presents an overview of the distribution of selected categories throughout the collection.

\begin{figure}
	\centering
	\includegraphics[width=\textwidth]{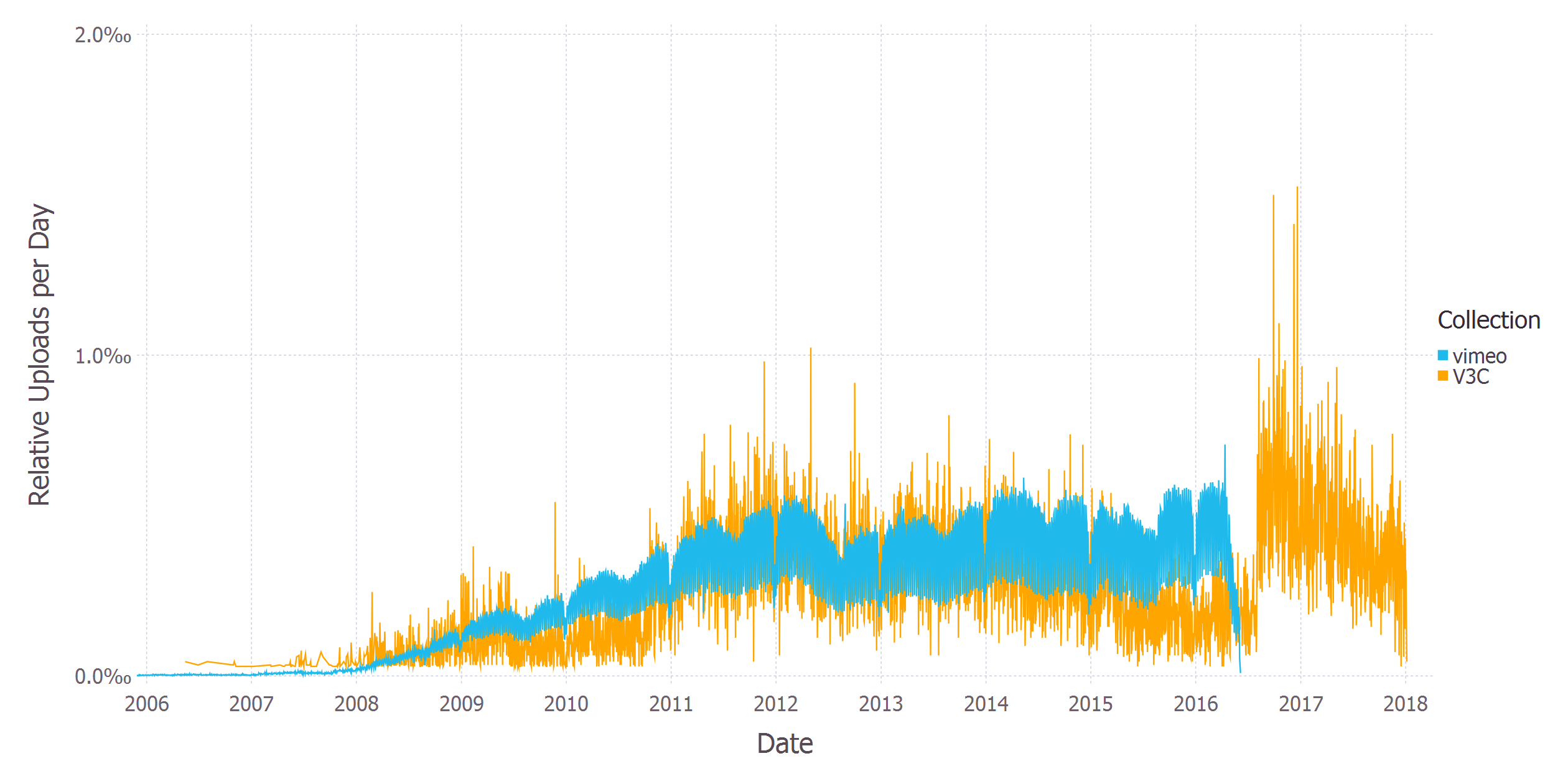}
	\caption[Daily uploads for the V3C]{Daily relative video uploads from the V3C and the Vimeo dataset}
	\label{fig:3:v3c-upload}
\end{figure}

The age distribution of the videos of the entire collection as determined by the upload date of the individual video is illustrated in Figure \ref{fig:3:v3c-upload}. It is shown in comparison to the distribution originally presented in \cite{rossetto2017web} for a large sample of Vimeo in general. The trace representing the V3C is less clean than the one for the Vimeo dataset due to the large difference in number of data points. It can however still be seen that both traces have a similar overall shape, at least for the parts of the plot where there is data available for both. Other than the Vimeo dataset from~\cite{rossetto2017web}, the collection of which was completed mid 2016, the V3C includes videos from as late as early 2018 which explains the difference in shape towards the right side of the plot.

\begin{figure}
	\centering
	\includegraphics[width=\textwidth]{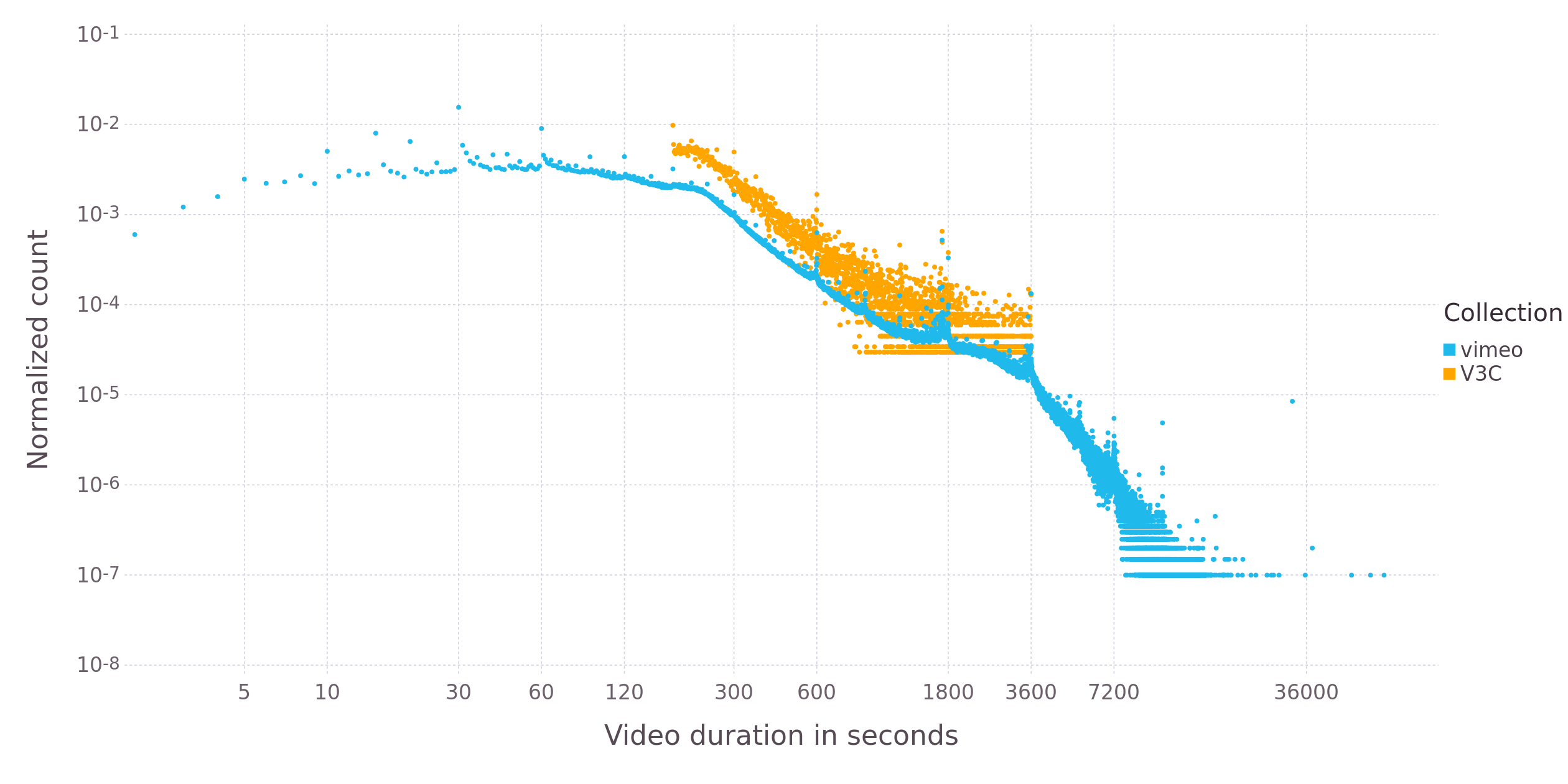}
	\caption{Scatter plot showing the duration of videos from the V3C and the Vimeo dataset}
	\label{fig:3:duration-scatter-v3c}
\end{figure}

\begin{figure}
	\centering
	\includegraphics[width=\textwidth]{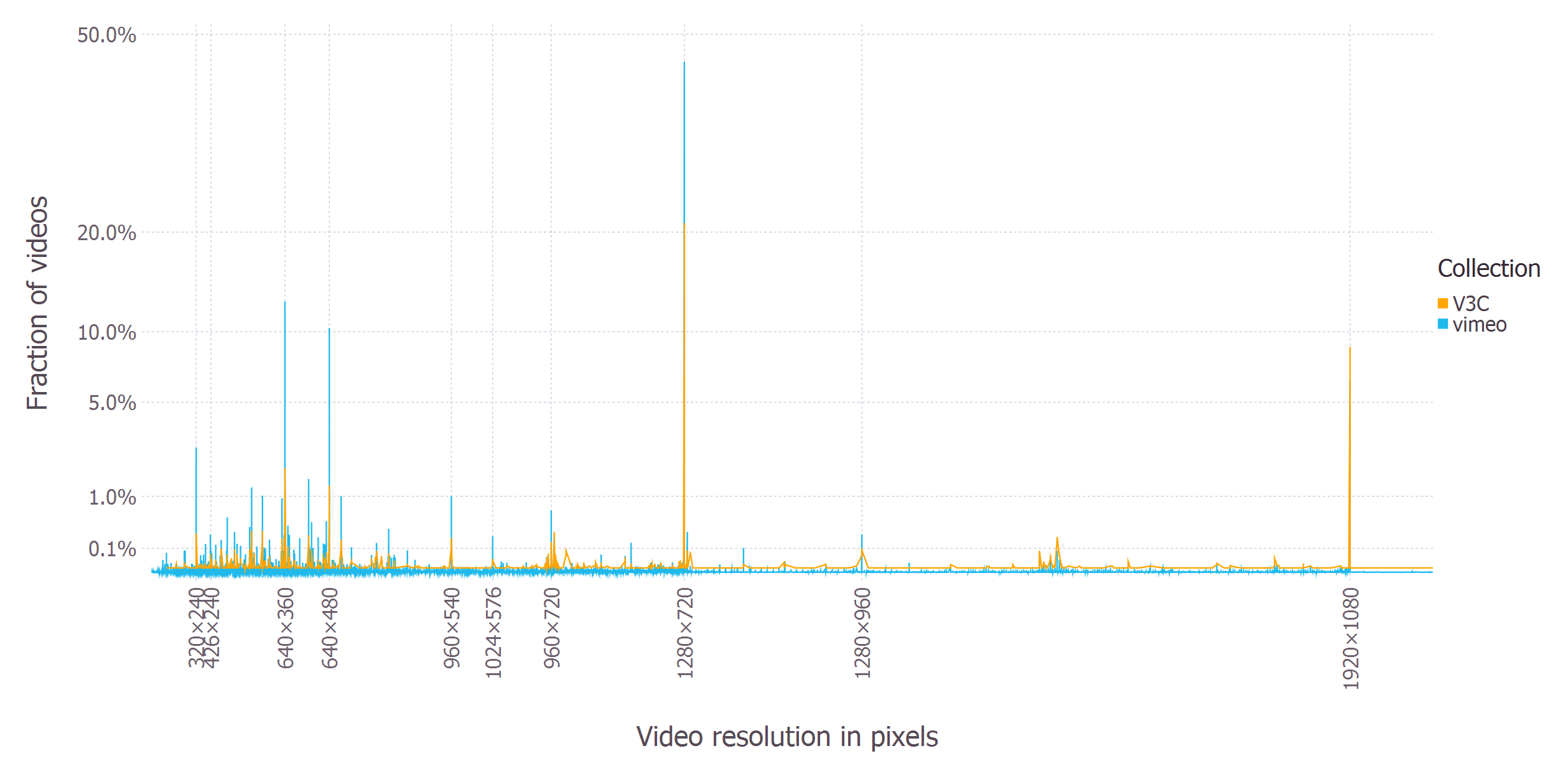}
	\caption{Distribution of video resolutions in the V3C}
	\label{fig:3:resolution-v3c}
\end{figure}

The distribution of video duration and resolution is shown in Figures \ref{fig:3:duration-scatter-v3c} and \ref{fig:3:resolution-v3c} respectively, again in comparison to the larger Vimeo distributions. It can be seen that wherever there were no additional restrictions, the properties of the V3C follow those of the overall Vimeo dataset rather closely. At least in terms of these three properties, the V3C can therefore be considered reasonably representative of the type of web video generally found on Vimeo.

%\textcolor{red}{G.A : Do we know how many different languages are included in the V3C collection? If so, it is interesting to have a table/graph showing the distribution}

An overview of the languages detected by the same method as employed in~\cite{rossetto2017web}, based on the title and description of the videos can be seen in Table~\ref{tab:languages}. It shows the top-10 languages for either the V3C or the dataset from \cite{rossetto2017web}. The column labeled `?' represents the instances where language detection did not yield any result. It can be seen that for the videos, the titles and descriptions of which were distinct enough for language detection, the distribution within the V3C is similar to the Vimeo dataset. No language analysis based on the audio data of the videos has been performed yet.

\begin{table}[b]
\centering
\caption{Overview of the detected languages in the video title and description of the V3C in percent}
	\label{tab:languages}
    \setlength{\tabcolsep}{0.45em}
\begin{tabular}{l|c|c|c|c|c|c|c|c|c|c|c|c|}
\cline{2-13}
 & ? & en & de & fr & it & es & cy & pl & nl & pt & ko & ru \\ \hline
\multicolumn{1}{|l|}{Vimeo} & 63.07 & 27.36 & 1.38 & 1.35 & 0.62 & 0.48 & 0.24 & 0.37 & 0.3 & 0.66 & 0.62 & 0.43 \\ \hline
\multicolumn{1}{|l|}{V3C} & 69.87 & 24.5 & 1.36 & 1.11 & 0.47 & 0.41 & 0.36 & 0.32 & 0.26 & 0.26 & 0 & 0 \\ \hline
\multicolumn{1}{|l|}{V3C1} & 68.52 & 25.34 & 1.65 & 1.23 & 0.54 & 0.64 & 0.31 & 0.29 & 0.28 & 0.25 & 0 & 0 \\ \hline
\multicolumn{1}{|l|}{V3C2} & 70.83 & 23.85 & 1.21 & 1.11 & 0.44 & 0.33 & 0.33 & 0.35 & 0.22 & 0.19 & 0 & 0 \\ \hline
\multicolumn{1}{|l|}{V3C3} & 69.94 & 24.63 & 1.3 & 1.04 & 0.45 & 0.33 & 0.41 & 0.32 & 0.29 & 0.26 & 0 & 0 \\ \hline
\end{tabular}
\end{table}

Table \ref{tab:categories} shows the categories and the number of videos per collection part which have been assigned to a particular category on Vimeo. Every video can be assigned to multiple categories, the numbers shown in the table do therefore not sum to the total number of videos. Despite the categories having a structure which implies a hierarchy, a video can be assigned to both a category and subcategory, but it does not have to. The large number of used categories shown in the table implies a wide range of content which can be found in the collection.

\begin{longtable}[c]{|l|r|r|r|}
\caption{Category assignment per video and collection part}
\label{tab:categories}\\
\hline
\multirow{2}{*}{Vimeo Category} & \multicolumn{3}{l|}{Number of videos} \\ \cline{2-4} 
 & V3C1 & V3C2 & V3C3 \\ \hline
\endfirsthead
\endhead
/categories/art & 660 & 891 & 1'010 \\ \hline
/categories/art/homesandliving/videos & 11 & 11 & 13 \\ \hline
/categories/art/personaltechdesign/videos & 15 & 17 & 14 \\ \hline
/categories/cameratechniques & 513 & 703 & 749 \\ \hline
/categories/cameratechniques/drones/videos & 156 & 191 & 204 \\ \hline
/categories/cameratechniques/macroandslomo/videos & 12 & 18 & 14 \\ \hline
/categories/cameratechniques/timelapse/videos & 161 & 252 & 281 \\ \hline
/categories/comedy & 252 & 315 & 388 \\ \hline
/categories/comedy/comicnarrative/videos & 74 & 69 & 86 \\ \hline
/categories/documentary & 1'396 & 1'787 & 2'086 \\ \hline
/categories/documentary/artsandcraft/videos & 54 & 82 & 99 \\ \hline
/categories/documentary/cultureandtech/videos & 78 & 124 & 117 \\ \hline
/categories/documentary/nature/videos & 155 & 191 & 191 \\ \hline
/categories/documentary/people/videos & 206 & 272 & 342 \\ \hline
/categories/documentary/sportsdocumentary/videos & 17 & 32 & 34 \\ \hline
/categories/fashion & 166 & 226 & 255 \\ \hline
/categories/fashion/fashionprofiles/videos & 8 & 5 & 12 \\ \hline
/categories/food & 87 & 131 & 145 \\ \hline
/categories/food/profiles/videos & 15 & 26 & 24 \\ \hline
/categories/hd/canon/videos & 894 & 1'122 & 1'328 \\ \hline
/categories/hd/dslr/videos & 438 & 528 & 660 \\ \hline
/categories/hd/pockethd/videos & 2 & 4 & 4 \\ \hline
/categories/hd/red/videos & 16 & 27 & 25 \\ \hline
/categories/hd/slowmotion/videos & 23 & 36 & 45 \\ \hline
/categories/instructionals & 283 & 389 & 402 \\ \hline
/categories/instructionals/healthandfitness/videos & 38 & 52 & 61 \\ \hline
/categories/instructionals/martialarts/videos & 8 & 11 & 20 \\ \hline
/categories/instructionals/outdoorskills/videos & 4 & 6 & 6 \\ \hline
/categories/journalism & 991 & 1'209 & 1'544 \\ \hline
/categories/journalism/nonprofit/videos & 67 & 80 & 116 \\ \hline
/categories/journalism/politics/videos & 131 & 155 & 182 \\ \hline
/categories/journalism/startups/videos & 8 & 12 & 18 \\ \hline
/categories/journalism/videojournalism/videos & 182 & 226 & 305 \\ \hline
/categories/music & 1'066 & 1'347 & 1'568 \\ \hline
/categories/music/musicdocumentary/videos & 50 & 52 & 77 \\ \hline
/categories/narrative & 2'114 & 2'614 & 3'014 \\ \hline
/categories/narrative/comedicfilm/videos & 66 & 90 & 111 \\ \hline
/categories/narrative/drama/videos & 60 & 73 & 95 \\ \hline
/categories/narrative/horror/videos & 30 & 38 & 34 \\ \hline
/categories/narrative/lyrical/videos & 2 & 11 & 12 \\ \hline
/categories/narrative/musical/videos & 3 & 3 & 7 \\ \hline
/categories/narrative/romance/videos & 22 & 34 & 25 \\ \hline
/categories/narrative/scifi/videos & 19 & 14 & 17 \\ \hline
/categories/nature-toplevel-modonly & 0 & 0 & 1 \\ \hline
/categories/personal & 916 & 1'200 & 1'378 \\ \hline
/categories/personal/cameo/videos & 6 & 8 & 5 \\ \hline
/categories/personal/stories/videos & 158 & 221 & 246 \\ \hline
/categories/productsandequipment/cameras/videos & 13 & 25 & 27 \\ \hline
/categories/productsandequipment/editingproducts/videos & 49 & 74 & 86 \\ \hline
/categories/productsandequipment/lighting/videos & 12 & 13 & 25 \\ \hline
/categories/productsandequipment/producttutorials/videos & 13 & 9 & 10 \\ \hline
/categories/sports & 1'487 & 2'036 & 2'213 \\ \hline
/categories/sports/bikes/videos & 152 & 211 & 196 \\ \hline
/categories/sports/everythingelse/videos & 39 & 43 & 52 \\ \hline
/categories/sports/outdoorsports/videos & 392 & 522 & 604 \\ \hline
/categories/sports/skate/videos & 147 & 227 & 207 \\ \hline
/categories/sports/sky/videos & 84 & 132 & 172 \\ \hline
/categories/sports/snow/videos & 75 & 96 & 117 \\ \hline
/categories/sports/surf/videos & 76 & 104 & 110 \\ \hline
/categories/technology & 0 & 0 & 1 \\ \hline
/categories/technology/installations/videos & 3 & 8 & 9 \\ \hline
/categories/technology/personaltech/videos & 5 & 2 & 6 \\ \hline
/categories/technology/software/videos & 2 & 9 & 12 \\ \hline
/categories/technology/techdocs/videos & 16 & 21 & 18 \\ \hline
/categories/travel & 1'893 & 2'450 & 2'803 \\ \hline
/categories/travel/africa/videos & 22 & 15 & 24 \\ \hline
/categories/travel/antarctica/videos & 1 & 2 & 4 \\ \hline
/categories/travel/asia/videos & 55 & 82 & 73 \\ \hline
/categories/travel/australasia/videos & 7 & 12 & 10 \\ \hline
/categories/travel/europe/videos & 94 & 124 & 120 \\ \hline
/categories/travel/northamerica/videos & 44 & 53 & 56 \\ \hline
/categories/travel/southamerica/videos & 13 & 21 & 15 \\ \hline
/categories/travel/space/videos & 5 & 6 & 10 \\ \hline
/categories/videoschool & 1 & 0 & 0 \\ \hline
\end{longtable}

\subsection{Possible Uses}

Due to the large diversity of video content contained within the collection, it can be useful for video-related applications in multiple areas. The large number of different video resolutions -- and to a lesser extent frame-rates -- makes this dataset interesting for video transport and storage applications such as the development of novel encoding schemes, streaming mechanisms or error-correction techniques.

Its large variety in visual content makes this dataset also interesting for various machine learning and computer vision applications.

Finally, the collection has applications in the area of video analysis, retrieval and exploration. For example,
we can imagine four possible application areas in the video retrieval space. First, video tagging or high-level feature detection where the goal is given a video segment or shot, the system should output all the relevant tags and visual concepts that are in this video. Such a task is very fundamental to any video search engine that tries to match users search queries with video dataset to retrieve the most relevant results. Second, ad-hoc video search where a system takes as input a user text query as a natural language sentence and returns the most relevant set of videos that satisfies the information need in the query. Such a task is also necessary for any search system that deals with real users where it has to understand the user query and intention before retrieving the set of results that matches the text query. Third, trying to find a video or a video segment which one believes to have seen but the name of which one does not recall is often called “known item search”. Queries are created based on some knowledge of the collection such that there is a high probability that there is only
one video or video segment that satisfies the search. Fourth, the application of video captioning or description in recent years gained a lot of attention. Here the idea is how can a system describe a video segment in a textual form that contains all the important facets such as `who', `what', `where', `when' so essentially textual summary of the video. As the V3C collection includes a master shot boundary splitting a whole video into smaller shots, the video captioning task can be run on those small video shots as currently the state of the art can not handle longer videos and give a logical and human readable description for the whole video in textual form.

\subsection{Availability}
We are planning to launch and make available this collection at the 2019 TRECVid video retrieval benchmark where different research groups participate in one or more tracks. In addition, the collection will be shared at the Interactive Video Browser Showdown (VBS)~\cite{cobarzan2017interactive} which collaborates with TRECVid organizing the Video Ad-hoc Search track. The collection will be available to the benchmark participants as well as the public for download. After the annual benchmark cycle is concluded, we will also provide the ground truth judgments and queries/topics for the tasks that used the V3C collection so that research groups can reuse the dataset in their local experiments and reproduce results.

%\todo[inline]{should we even put this directly into the paper?}
%\textcolor{red}{G.A : See if what I added make sense. So far I avoided talking about TRECVID explicitly. We can add this after it gets accepted, before camera ready version}
%L.R: Yes, I think the part is clear.

\section{Conclusions}
\label{sec:conclusion}
In this paper, we introduced the Vimeo Creative Commons Collection (V3C). It is comprised of roughly 3'800 hours of creative commons video obtained from the web video platform Vimeo and is augmented with technical and semantic metadata as well as shot boundary information and accompanying keyframes. V3C is subdivided into three partitions with increasing length from roughly 1'000 hours up to 1'500 hours so that the collection can be used for at least three consecutive years in a video search benchmark with increasing complexity. Information on where to download the V3C collection and/or its partitions will be made available together with the publication of the video search benchmark challenges.

% \todo[inline]{what else?}

\section*{Acknowledgements}
This  work  was  partly  supported  by  the  Swiss  National Science Foundation, project IMOTION (20CH21\_151571).

\emph{Disclaimer: Certain commercial entities, equipment, or
materials may be identified in this document in order to describe an
experimental procedure or concept adequately. Such identification is
not intended to imply recommendation or endorsement by the National
Institute of Standards and Technology, nor is it intended to imply 
that the entities, materials, or equipment are necessarily the best 
available for the purpose.}

\bibliographystyle{plain}
\bibliography{bibliography}

\end{document}